\documentclass[sigconf,usenames,dvipsnames]{acmart}
\usepackage{courier}
\usepackage{url}
\usepackage{balance}
\usepackage{enumitem}
\usepackage{caption}	
\usepackage{multirow}
\usepackage{multicol}
\usepackage{amsmath}
\usepackage{float}
\usepackage[normalem]{ulem}
\usepackage{xspace}
\usepackage{listings}
\usepackage{subcaption}
\usepackage{hyperref}

\newcommand{\ie}{\textit{i.e.,}\xspace}
\newcommand{\eg}{\textit{e.g.,}\xspace}

\newcommand{\mplus}{{\sc MDroid+}\xspace}

\lstdefinestyle{interfaces}{
  float,
  floatplacement=tb
}

\include{macro}
\begin{document}

\title{MDroid+: A Mutation Testing Framework for Android}
\author{Kevin Moran}
\affiliation{College of William and Mary \\United States}
\author{Michele Tufano}
\affiliation{College of William and Mary \\United States}
\author{Carlos Bernal-C\'ardenas}
\affiliation{College of William and Mary \\United States}
\author{Mario Linares-V\'{a}squez}
\affiliation{Universidad de los Andes \\ Colombia}
\author{Gabriele Bavota}
\affiliation{Universit\`{a} della Svizzera italiana \\ Switzerland}
\author{Christopher Vendome}
\affiliation{College of William and Mary \\United States}
\author{Massimiliano Di Penta}
\affiliation{University of Sannio \\ Italy}
\author{Denys Poshyvanyk}
\affiliation{College of William and Mary \\United States}

\renewcommand{\shortauthors}{K. Moran, M. Tufano, C. Bernal Cardenas, M. Linares Vasquez et al.}

\begin{abstract}

Mutation testing has shown great promise in assessing the effectiveness of test suites while exhibiting additional applications to test-case generation, selection, and prioritization. Traditional mutation testing typically utilizes a set of simple language specific source code transformations, called operators, to introduce faults.  However, empirical studies have shown that for mutation testing to be most effective, these simple operators must be augmented with operators specific to the domain of the software under test.  One challenging software domain for the application of mutation testing is that of mobile apps.  While mobile devices and accompanying apps have become a mainstay of modern computing, the frameworks and patterns utilized in their development make testing and verification particularly difficult. As a step toward helping to measure and ensure the effectiveness of mobile testing practices, we introduce \mplus, an automated framework for mutation testing of Android apps. \mplus includes 38 mutation operators from ten empirically derived types of Android faults and has been applied to generate over 8,000 mutants for more than 50 apps. 
 
 \noindent \textbf{Video URL: \url{https://youtu.be/yzE5_-zN5GA}}

\end{abstract}



%ACM Copyright Information Provided by the ACM rights management system

\copyrightyear{2018} 
\acmYear{2018} 
\setcopyright{rightsretained} 
\acmConference[ICSE '18 Companion]{40th International Conference on Software Engineering }{May 27-June 3, 2018}{Gothenburg, Sweden}
\acmBooktitle{ICSE '18 Companion: 40th International Conference on Software Engineering , May 27-June 3, 2018, Gothenburg, Sweden}\acmDOI{10.1145/3183440.3183492}
\acmISBN{978-1-4503-5663-3/18/05}

\maketitle

\section{Introduction}
\label{sec:introduction}

	Mobile devices have become a mainstay of the modern computing landscape. The increasing popularity of these devices is driven primarily by a rich ecosystem of applications, commonly referred to as ``apps".  Strong user demand for mobile apps has driven increased competition on storefronts such as Google Play~\cite{GP} and Apple's App Store~\cite{AppleStore}.  These competitive marketplaces necessitate that mobile developers devote time to extensively test and validate their apps to ensure a positive user experience, as new app releases have been shown to affect users' perception via marketplace ratings \cite{Martin:FSE16}.  Android is currently the most popular OS in the world~\cite{android-popularity}, and hence garners a large amount of interest from developers. 
	
	However, effectively testing Android apps is difficult, as developers face constraints specific to the domain of mobile software including change-prone APIs~\cite{Linares-Vasquez:FSE13,Bavota:TSE15}, platform fragmentation~\cite{Han:WCRE12}, and a lack of automated testing tools that meet developer needs~\cite{Linares:ICSME17:I}.  Furthermore, when devising testing strategies or writing test cases, their effectiveness needs to be evaluated.	To this aim, mutation analysis has been defined as a process for evaluating the efficacy of a test suite and functions by injecting small changes, meant to represent common bugs, into a ``correct'' program and then evaluating the number of injected faults uncovered by the test suite~\cite{Hamlet:TSE,DeMillo:Computer}.  The higher the ratio of uncovered bugs, or \textit{mutants}, the more effective a test suite is said to be.  However, mutation analysis typically relies on a concept called the \textit{coupling effect} that states that
\textit{``complex faults are coupled to simple faults in such a way that a test data set that detects all simple faults in a program will detect most complex faults''}~\cite{Offutt2001}.  The validity of this effect has come under scrutiny from the software testing research community \cite{DaranT96,Andrews:ICSE05,AndrewsBLN06,Just:FSE14,Luo:FSE16}, and while studies have generally indicated a correlation between mutants and real faults, they also point out that non-negligible portion of real faults do not effectively map to mutants~\cite{Just:FSE14}. The number of simple mutants mapping to real faults in the mobile domain is likely to be lower, due to heavy usage of APIs and frameworks that enable the varied feature sets of mobile devices.  Thus, in order to develop an effective mutation testing tool for Android, one must take into account the software domain, and model operators according to faults that naturally occur in mobile software development.
	
	In this paper, we describe \mplus, a mutation testing framework for Android apps that aims to support developers in writing mobile tests.  \mplus includes 38 Android and Java specific mutation operators that were designed according to an empirically derived taxonomy of common, naturally occurring faults in Android applications \cite{Linares-Vasquez:FSE17}. The main contributions of \mplus can be summarized as follows:

\begin{itemize}[leftmargin=*]
	\item A set of empirically derived mutation operators designed to simulate commonly observed faults in Android apps;
	\item An efficient, automated methodology for deriving a Potential Fault Profile (PFP) from subject apps, dictating possible locations for mutant injection;
	\item A process for applying operator transformations to locations dictated by the PFP to create mutant apps;
 	\item Built-in extensibility that facilitates new operator definitions.
\end{itemize}
\vspace{-0.3cm}
\section{Approach}
\label{sec:approach}

\begin{figure}[t]
  \centering
\vspace{-0.2cm}
  \includegraphics[width=\linewidth]{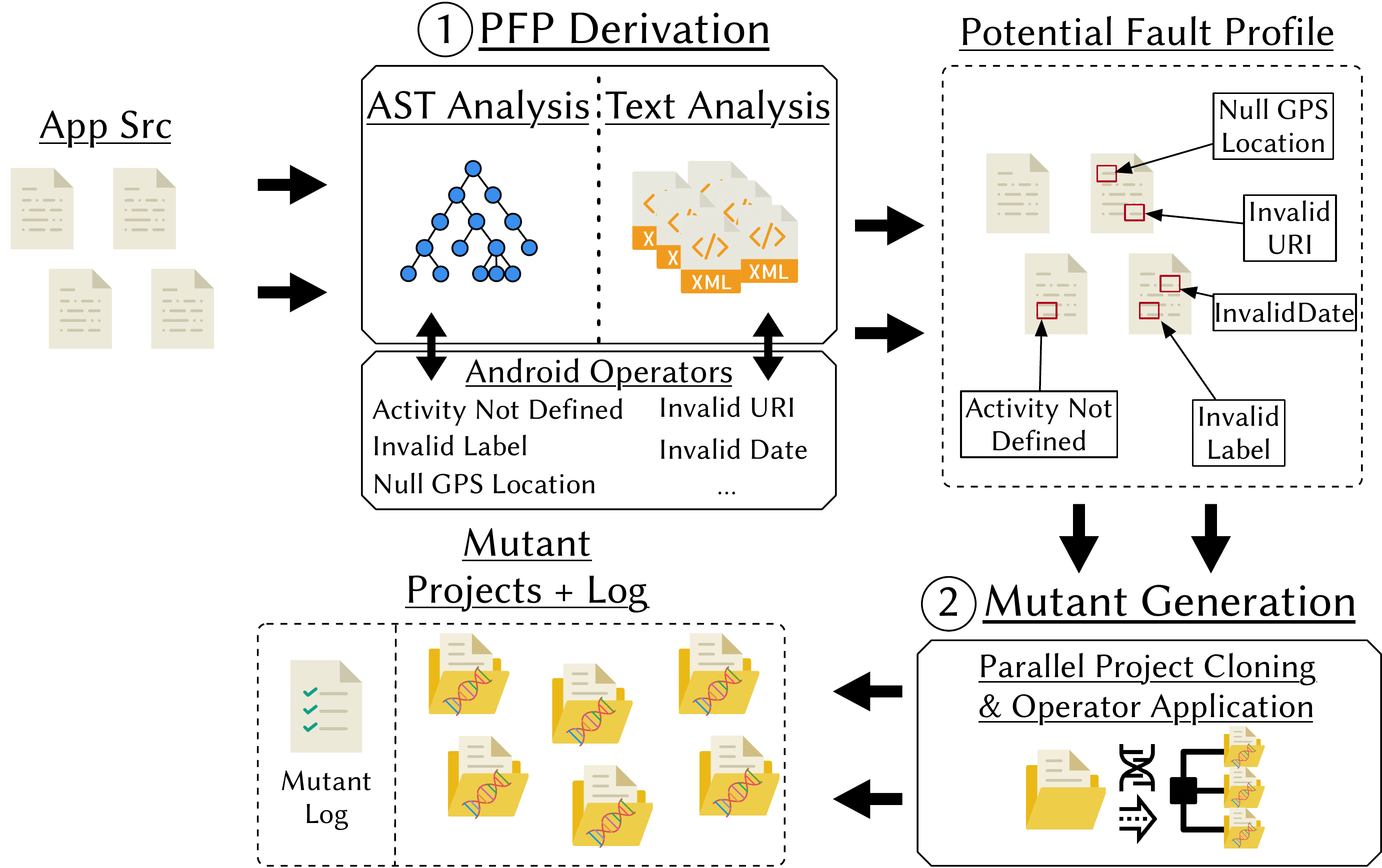}
\vspace{-0.7cm}
  \caption{Overview tool Workflow for \mplus}
%\vspace{-0.2cm}
  \label{fig:framework-overview}
\end{figure}

	In order to ensure that \mplus is an effective, practical, and flexible/extensible tool for mutation testing, it takes into account the following design considerations: (i) an empirically derived set of mutation operators; (ii) a design embracing the open/closed principle (\ie open to extension, closed to modification); (iii) visitor and factory design patterns for deriving the Potential Failure Profile (PFP) and applying operators, (iv) parallel computation for efficient mutant seeding.  \mplus is written in Java and available as an open source project \cite{replication}. In the following sections, we describe \mplus according to its workflow described in Figure \ref{fig:framework-overview}.

\vspace{-0.2cm}
\subsection{Implemented Operators}
\label{subsec:operators}

As mentioned in Section \ref{sec:introduction}, in order to create a mutation testing framework for Android, mutation operators must be defined according to naturally occurring faults to ensure a strong coupling between operators and faults likely to befall an Android project.  To this end, the authors undertook an extensive empirical study following a procedure inspired by open coding~\cite{Miles2013}. More specifically, the authors analyzed 2,007 documents, assigning a label describing an observed bug, from the following sources: (i) bug reports of open source Android apps, (ii) bug-fixing commits of open source Android apps, (iii) Stack-Overflow discussions, (iv) exception hierarchies of Android APIs, (v) bugs described in previous studies, and (vi) reviews from Google Play.  The open-coding procedure was supported by a custom designed web-application, where each document was tagged by at least two authors, and labeling conflicts were resolved. For the complete methodology, we refer readers to our previous technical paper \cite{Linares-Vasquez:FSE17}.

	The study described above resulted in a taxonomy of faults for Android apps that spans 14 different categories. We implemented 38 operators corresponding to ten different categories, excluding certain categories of operators such as hardware configurations, as they do not map well to the process of mutation testing.  A detailed list and description of all implemented operators is available on the \mplus website~\cite{replication}.

\vspace{-0.2cm}
\subsection{Derivation of the Potential Fault Profile}
\label{subsec:pfp-derivation}

In the context of \mplus, we define a PFP that stipulates locations in analyzed apps---which can correspond to source code statements, XML tags, or locations in other resource files---that coincide with potential fault locations, given the empirically derived taxonomy~\cite{Linares-Vasquez:FSE17}. Consequently, the PFP stipulates the locations where our defined mutation operators can be applied.  To extract the PFP, \mplus statically analyzes a target mobile app, looking for locations where operators can be implemented. The locations are detected automatically by parsing XML files in the case of resources or through Abstract Syntax Tree (AST) based analysis for Java code.  

	For implemented operators targeting one of the Android resource XML files (\ie archives in the \texttt{\small /res} folder), the structure of each XML file is analyzed and a pattern matching process for different attributes within the XML is used.  However, for operators that are applied to the Java source code, a two-phase AST-based and text-based analysis is utilized that is capable of identifying the location of target API calls.  The identification of API call sites is implemented utilizing the visitor design pattern, allowing for extensible, decoupled operations to be performed on the AST of a target app.  This helps to ensure that \mplus is capable of supporting additional operators in the future that may require more advanced AST analysis. After the AST-based location of specific API calls, a fine-grained, text-based pattern matching is performed on identified API calls to derive the precise textual location where the mutation operator transformation will be applied. The end result of the PFP derivation process is a list stipulating all potential injection points in code of the Android-specific mutation operators.

\vspace{-0.2cm}
\subsection{Mutant Creation}
\label{subsec:mutant-creation}

	Given an automatically derived PFP for an app and the catalog of Android-specific operators, \mplus generates a mutant for each location in the PFP. This process is performed using text or AST manipulation rules specific to each implemented operator.  Thus, for each location related to an operator, the text/AST  transformation is applied to the specified location in either the code or \texttt{\small .xml} file.  

	Due to the event driven nature of Android applications, testing is generally performed at the GUI-level and is centered around app use-cases \cite{Linares:ICSME17:I}.  Therefore, in order to operationalize \mplus to fit in a typical testing workflow, the mutated applications must be compilable to an \texttt{\small .apk} file that can be run on an emulator or real device.  Thus, during the mutant creation process, \mplus creates a project-level clone of a target and applies a single mutation to a specified location in the cloned project, resulting in one mutant project for each seeded instance of a mutation operator.  It is important to note that the cost of applying the transformations to the cloned projects is trivial in practice, and a majority of the execution time cost is related to the cloning of the app project, and is thus I/O bound.  To ensure the cloning and mutant generation process proceeds as efficiently as possible, \mplus implements an option to parallelize the process, utilizing the multi-core architecture of most modern hardware.  The end result of the mutation creation process is the set of cloned mutant projects and a log that delineates the operator and location of injection for each created mutant.  Currently, we do not offer a universal interface for the compilation of mutants, as the parameters and build systems used by Android apps can vary dramatically.  However, the compilation process can be easily scripted using the CLI support offered by \texttt{\small ant}, \texttt{\small gradle}, and the Android SDK\footnote{\url{https://developer.android.com/studio/build/building-cmdline.html}}.

\vspace{-0.2cm}
\subsection{Tool Usage and Extensibility}
\label{subsec:tool-usage}

	\mplus is implemented as a Java command-line utility in which the user can select the specific set of mutation operators to be applied during mutant generation, as well as an option for enabling multi-threading.  A {\sc Readme} and user guide can be found on the project repository, which is accessible from the tool website \cite{replication}.

	Given that the Android platform is prone to rapid evolution~\cite{Linares-Vasquez:FSE13,Bavota:TSE15}, it is important that \mplus allows for easy modification/extension of the operators list, in order to keep pace with rapid evolution.  To add a new operator to \mplus, there are two major components that must be implemented: (i) an operator locator/detector, and (ii) the operator transformation rules.  If a proposed operator applies to the source code of an application, then either the target API-call of the operator must be added to an existing AST visitor or a new visitor to identify the AST pattern required by the operator must be defined.  Next, an operator specific pattern locator must be implemented to derive the precise operator location after API-calls in question have been detected via AST-analysis. To allow for streamlined extensibility, \mplus offers a generic \texttt{\small Locator} interface that includes a \texttt{\small findExactLocations()} method that can be implemented to support additional operators. An example of this interface implemented for the \texttt{\small BuggyGUIListener} operator is given in Listing 1. This locator simply returns the start and end of the API-call location passed in the \texttt{\small MutationLocation} object so that it can be manipulated later. Note, that this is a simple example, and most of \mplus's pre-defined operators need to parse particular properties or patterns from identified API calls. If the proposed operator applies to the resource \texttt{\small .xml}, \mplus implements a \texttt{TextBasedDetector} abstract class that can be extended to account for matching the patterns of xml attributes.

\vspace{-0.2cm}
\begin{lstlisting}[language=Java, caption={Example of Operator Locator}]
public class BuggyGUIListenerLocator implements Locator {

   private void findExactLocation(MutationLocation loc) {
		//Fix start column
		loc.setStartColumn(loc.getStartColumn()+1);
		//Build exact mutation location
		loc.setEndColumn(loc.getStartColumn()+loc.getLength());
   }
}
\end{lstlisting}

	 Once the precise locator/detector for a proposed mutation operator has been implemented, the actual transformation rule must be delineated so that it can be applied to any detected injection points during the \mplus analysis.  To facilitate this process, \mplus includes a \texttt{\small MutationOperator} interface and implements the factory design pattern for managing and instantiating operators.  To add an additional transformation rule for a desired operator, the \texttt{\small performMutation()} method of the interface must be implemented according to the \texttt{\small MutationLocation}.  An example of the \texttt{NullInp utStream} operator that sets an \texttt{inputStream} to null before it is closed is shown in Listing 2.

\vspace{-0.3cm}
\begin{lstlisting}[language=Java, caption={Example of Operator Definition}]
public class NullStream  implements MutationOperator{
	@Override
	public boolean performMutation(MutationLocation location) {
		ObjectMutationLocation mLocation = (ObjectMutationLocation) location;
		List<String> newLines = new ArrayList<String>();
		List<String> lines = FileHelper.readLines(location.getFilePath());
		for(int i=0; i < lines.size(); i++){
			String currLine = lines.get(i);
			//Null object
			if(i == location.getLine()){
				String newLine = mLocation.getObject() + " = null;";
				newLines.add(newLine);	
			}
			newLines.add(currLine);
		}
		FileHelper.writeLines(location.getFilePath(), newLines);
		return true;
	}
}
\end{lstlisting}
\vspace{-0.4cm}

\section{Evaluation}
\label{sec:evaluation}

%---------------------------------
\subsection{Study Context}

To evaluate \mplus, we conducted a study with the following \emph{goals}: (i) understand and compare the \textit{applicability} of \mplus and other currently available mutation testing tools to Android apps; (ii) to understand the \textit{underlying reasons} for non-compilable or non-executable (\ie \textit{trivial}) mutants. To accomplish this, we compare \mplus with two popular open source mutation testing tools (Major~\cite{Just:ISSTA14} and PIT~\cite{PIT}), which we adapted to be applicable to Android apps, and with one context-specific mutation testing tool for Android called muDroid \cite{Deng:ICSTW15}.  We chose these tools because of their diversity (in terms of functionality and mutation operators), their compatibility with Java, and their representativeness of tools working at different representation levels: source code, Java bytecode, and smali bytecode (\ie Android-specific bytecode representation).  The modifications to the PIT and Major tools can be accessed using our replication package \cite{replication}. To compare the applicability of each mutation tool, we used the Androtest suite of apps \cite{Shauvik:2015}, which includes 68 Android apps from 18 Google Play categories. The mutation testing tools exhibited issues in 13 of the considered 68 apps, \ie the mutant-injection process rendered them non-compilable. Thus, in the end, we considered 55 subject apps in our study. For more information see our replication package~\cite{replication}.

To quantitatively assess the applicability and effectiveness of the considered mutation tools to Android apps, we defined three metrics: \textit{Total Number of Generated Mutants} \textbf{(TNGM)}, \textit{Non-Compilable Mutants} \textbf{(NCM)}, and \textit{Trivial Mutants} \textbf{(TM)}.  In this paper, we consider \textit{non-compilable mutants} as those that are syntactically incorrect and cause compilation/assembly errors, and \textit{trivial mutants} as those that are killed arbitrarily by most test cases (\eg crashing on launch).  The trivial mutant study was supported by a large-scale dynamic analysis framework \cite{Linares-Vasquez:FSE17}.

%---------------------------------
\subsection{Study Results}

\begin{figure}[t]
	\vspace{-0.6cm}
	\begin{subfigure}[b]{0.225\textwidth}
		\subcaption{\%Non-Compilable Mutants}
		\vspace{-0.35cm}
		\includegraphics[width=\textwidth]{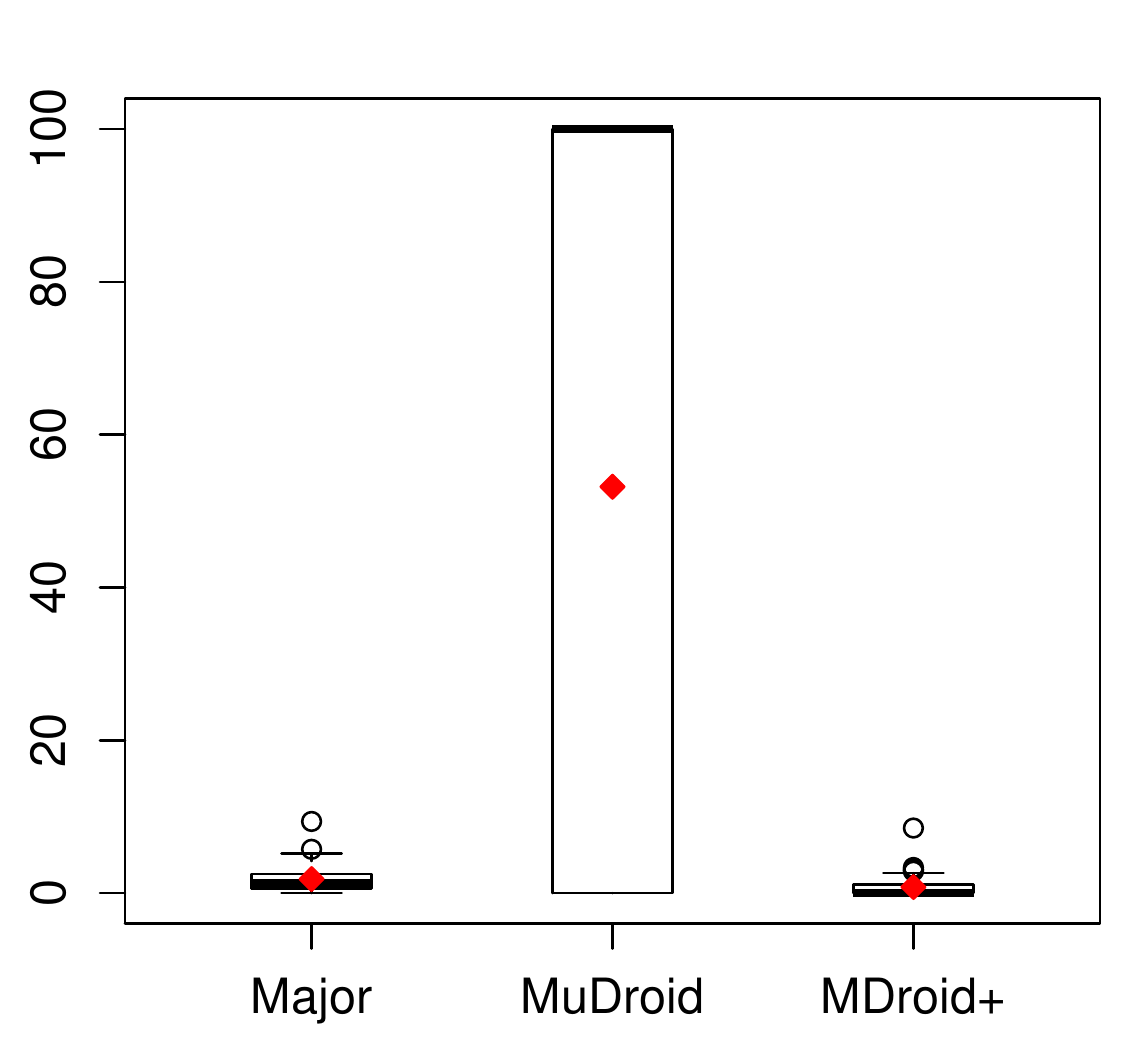}
		\label{fig:stillborn-mutants}
	\end{subfigure}
	\begin{subfigure}[b]{0.225\textwidth}
		\subcaption{\%Trivial Mutants}
		\vspace{-0.4cm}
		\includegraphics[width=\textwidth]{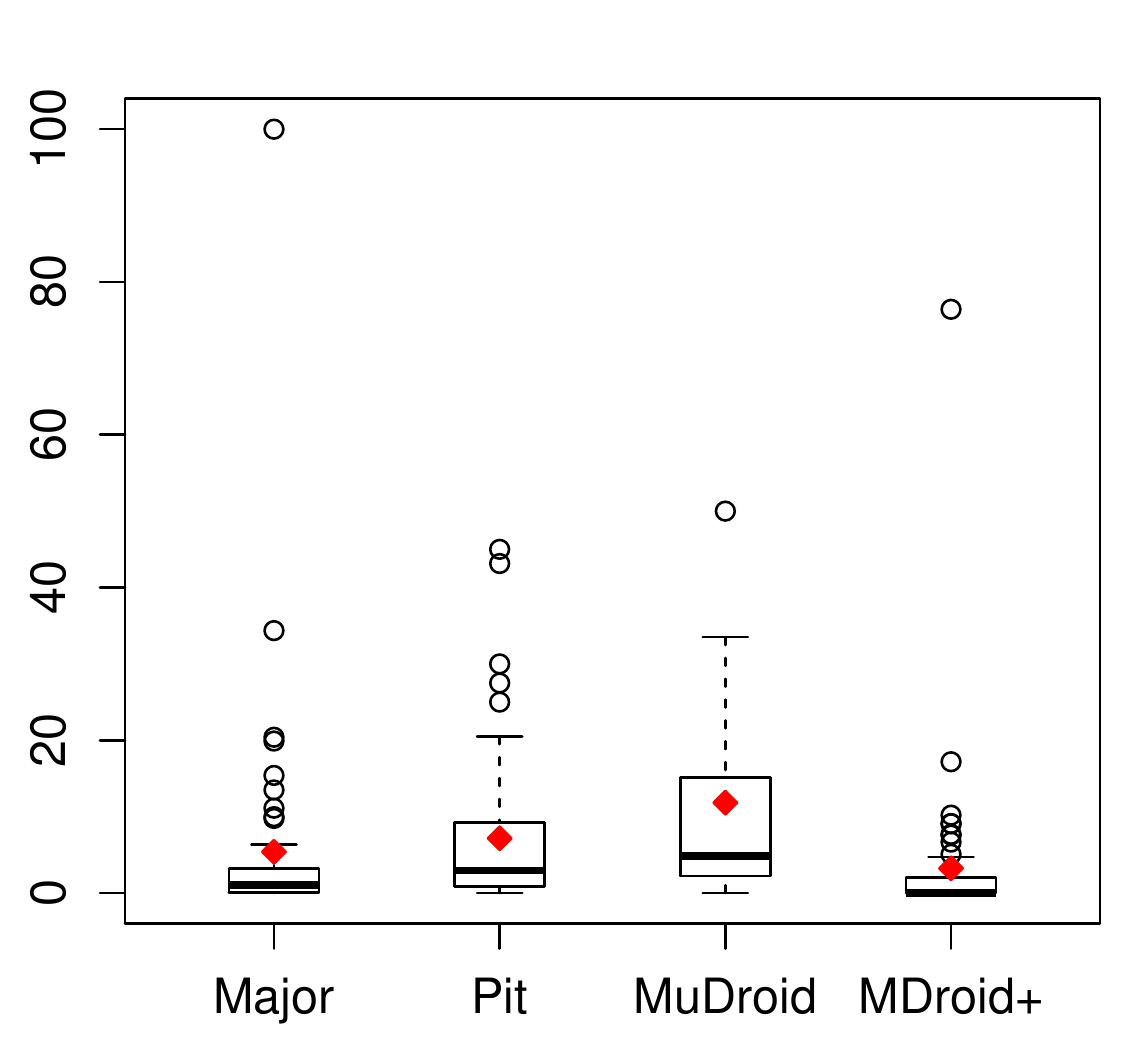}
		\label{fig:trivial-mutants}
	\end{subfigure}
	\begin{subfigure}[b]{0.43\textwidth}
		\vspace{-0.6cm}
		\subcaption{\# of Mutant Generated per App}
		\label{fig:gen-mutants}
		\vspace{-0.4cm}
		\includegraphics[width=\textwidth]{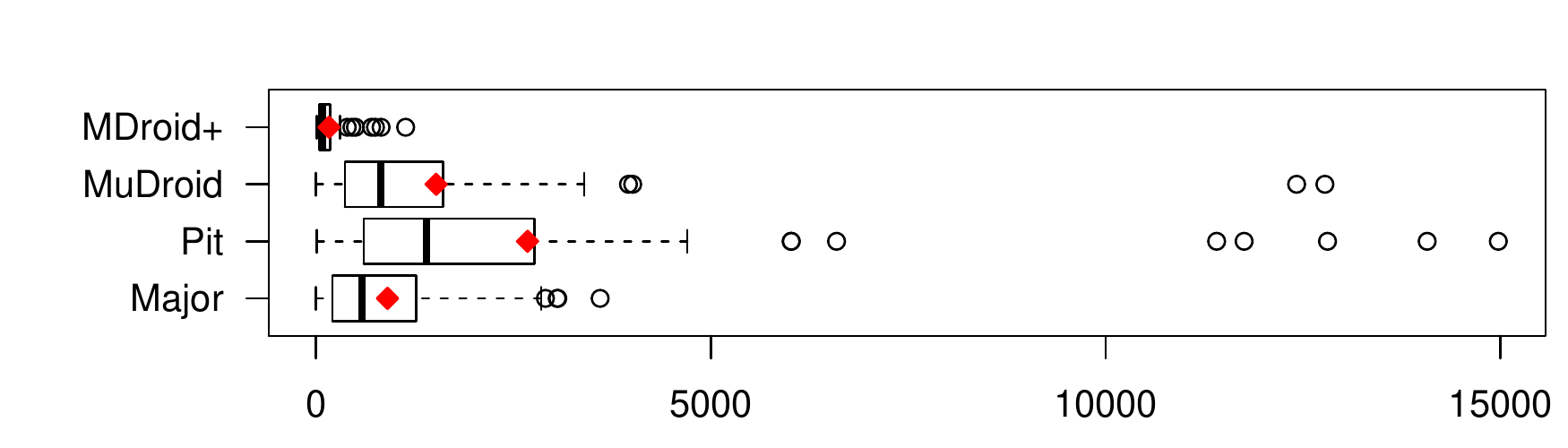}
	\end{subfigure}
	\vspace{-0.5cm}
	\caption{Non-compilable \& trivial mutants per app.}\label{fig:results}
	%\vspace{-0.1cm}
\end{figure}

	Figure \ref{fig:results} reports the results of (i) the percentage of non-compilable mutants (NCM), (ii) the percentage of trivial mutants (TM), and (iii) the total number of generated mutants per app. On average, 167, 904, 2.6k+, and 1.5k+ mutants were generated by \mplus, Major, PIT, and muDroid, respectively for each app.  \mplus had an average runtime of 19 seconds per app on server-class hardware. While the number of mutants generated is high for other tools, this is mostly due to the fact they implement far more general operators with a higher number of potential injection points. However, \mplus's operators are more specific, and coupled to features that may or may not be implemented by a given app, resulting in fewer, but stronger, mutants. The average percentage of \textit{non-compilable mutants} (NCM) generated by \mplus, Major and muDroid over all the apps is 0.56\%, 1.8\%, and 53.9\%, respectively, while no NCM are generated by PIT due to its mutant assembly process (Figure~\ref{fig:stillborn-mutants}). Thus, \mplus achieves the lowest ratio of non-compilable mutants by a statistically significant margin (Wilcoxon paired signed rank test  $p$-value$<0.001$ for Major and muDroid -- adjusted with Holm's correction \cite{holm}, Cliff's $d$=0.59 - large for Major, and Cliff's $d$=0.35 - medium for muDroid), allowing for a mutation testing procedure more capable of evaluating the efficacy of a set of tests . The overall rate of NCM is very low for \mplus, and most instances pertain to edge cases requiring more robust static analyses.

	All four tools generated \textit{trivial mutants}, and the mean percentage of their distribution over all apps for \mplus, Major, PIT, and muDroid is 2.42\%, 5.4\%, 7.2\%, and 11.8\%, respectively (Figure~\ref{fig:trivial-mutants}).  \mplus generates significantly fewer TM than muDroid (Wilcoxon paired signed rank test adjusted $p$-value=0.04, Cliff's $d$=0.61 - large) and PIT (adjusted $p$-value=0.004, Cliff's $d$=0.49 - large), while there is no statistically significant difference with Major (adjusted $p$-value=0.11). While these percentages may appear small, the raw values show that TMs can comprise a large set of instances for tools that can generate thousands of mutants per app. For example, for the Translate app, 518 out of the 1,877 mutants generated by PIT were TM. For the same app, muDroid creates 348 TM out of the 1,038 it generates.  The biggest underlying reason for TM in \mplus results from null objects or references in the \texttt{\small MainActivity}, causing a crash on startup. Future improvements to the tool could avoid mutants seeded in components related to the MainActivity.

\section{Demo Remarks \& Future Work}
\label{sec:approach}
\vspace{-0.1cm}

In this demo, we have presented \mplus, a mutation testing tool for Android applications that supports 38 empirically derived mutation operators, automates the process of detecting potential mutant locations and generating mutants, and facilitates the addition of new operators and the maintenance of existing operators through an extensible architecture.  \mplus was evaluated against other popular mutation testing tools for the Java language and was shown to generate fewer non-compilable and trivial mutants.  In the future, we plan to add to the functionality of \mplus by allowing users to stipulate specific activities for mutant injection and supporting the increasingly popular Kotlin language.

\textbf{Acknowledgments.} Bavota was supported in part by the SNF project JITRA, No. 172479. We would like to thank Sebasti\'an Sierra and Camilo Escobar from Universidad de los Andes for their help testing \mplus.

\balance
\bibliographystyle{abbrv}

\bibliography{ms}
\end{document}